# TIME SERIES FORECASTING: A NONLINEAR DYNAMICS APPROACH

*Stefano Sello*
*Termo-Fluid Dynamics Research Center*
*Enel Research*
*Via Andrea Pisano, 120*
*56122 PISA - ITALY*
*e-mail: sello@pte.enel.it*

*ABSTRACT*

The problem of prediction of a given time series is examined on the basis of recent nonlinear dynamics theories. Particular attention is devoted to forecast the amplitude and phase of one of the most common solar indicator activity, the international monthly smoothed sunspot number. It is well known that the solar cycle is very difficult to predict due to the intrinsic complexity of the related time behaviour and to the lack of a successful quantitative theoretical model of the Sun magnetic cycle. Starting from a previous recent work, we checked the reliability and accuracy of a forecasting model based on concepts of nonlinear dynamical systems applied to experimental time series, such as embedding phase space, Lyapunov spectrum, chaotic behaviour. The model is based on a local hypothesis of the behaviour on the embedding space, utilizing an optimal number k of neighbour vectors to predict the future evolution of the current point with the set of characteristic parameters determined by several previous parametric computations. The performances of this method suggest its valuable insertion in the set of the called statistical-numerical prediction techniques, like Fourier analyses, curve fitting, neural networks, climatological, etc. The main task is to set up and to compare a promising numerical nonlinear prediction technique, essentially based on an inverse problem, with the most accurate predictive methods like the so-called "precursor methods" which appear now reasonably accurate in predicting "long-term" Sun activity, with particular reference to the "solar" precursor methods based on a solar dynamo theory.

*INTRODUCTION*

Solar activity forecasting is an important topic for various scientific and technological areas, like space activities related to operations of low-Earth orbiting satellites, electric power transmission lines, geophysical applications, high frequency radio communications. The particles and electromagnetic radiations flowing from solar activity outbursts are also important to long term climate variations and thus it is very important to know in advance the phase and amplitude of the next solar and geomagnetic cycles. Nevertheless, the solar cycle is very difficult to predict on the basis of time series of various proposed indicators, due to high frequency content, noise contamination, high dispersion level, high variability in phase and amplitude. This topic is also complicated by the lack of a quantitative theoretical model of the Sun magnetic cycle. Many attempts to predict the future behavior of the solar activity are well documented in the literature. Numerous techniques for forecasting are developed to predict accurately phase and amplitude of the future solar cycles, but with limited success. Depending on the nature of the prediction methods we can distinguish five classes: 1) Curve fitting, 2) Precursor, 3) Spectral, 4) Neural networks, 5) Climatology.

Apart from precursor methods, the main limitation is the short time interval of reliable extrapolations, as the case of the McNish-Lincoln curve fitting method. [1] In the climatological method we predict the behaviour of the future cycle by a weighted average from the past N cycles, based on the assumption of a some degree of correlation of the phenomenon. A recent multivariate stochastic approach inside this class of methods is documented in [2].

A modern class of solar activity prediction methods appears to reasonably successful in predicting "long range" behavior, the precursor methods. Precursor are early signs of the size of the future solar activity that manifest before the clear evidence of the next solar cycle. There are two kind of precursor methods: geomagnetic (Thompson, 1993) [3], and solar (Schatten, 1978,1993) [4]. The basic idea is that if these methods work they must be based upon solar physics, in particular a dynamo theory. The precursor methods invoke a solar dynamo mechanism, where the polar field in the descending phase and minimum is the sign of future developed toroidal fields within the sun that will drive the solar activity (Schatten, Pesnell,1993). The dynamo method was successfully tested with different solar cycles with a proper statistical approach and verified by a scientific panel supported by NOAA Space Enviroment Center and NASA Office of Space Science, (1996,1997) [5]. The panel recommendations for future solar activity studies was based on some criticisms about long term solar cycle prediction because the weak physical basis of such predictions and the limitations of the data used to define and extend solar and geophysical behaviour: *"prediction research should be supported and the scientific community encouraged to develop a fundamental understanding of the solar cycle that would provide the basis for physical rather than the present empirical prediction methods"*.

Although the dynamo method and in general the precursor methods, seems to work well, they might be affected by some severe drawbacks, like telescope drifts and secular drifts of non-magnetic solar wind parameters. However, as pointed out by the authors, we need a better scientific basis.

Thus, at present, the statistical-numerical approach, based on some reliable characterization and prediction of the complex time series behaviour, without any intermediate model, it still appears as a valuable technique to provide at least the basis for future physical prediction methods.

The international sunspot number is a index characterizing the level of solar activity and it is regurarly provided by the Sunspot Index Data Center of the Federation of Astronomical and Geophysical data analysis Services. [6] The predictions are confined to the so called smoothed monthly sunspot number, a particular filtered signal from the monthly sunspot number. Figure 1 shows the time series of the monthly sunspot number (blue line) and the related smoothed monthly sunspot number (red line) from SIDC for the last two solar cycles and for the current 23th cycle. (June 1999)

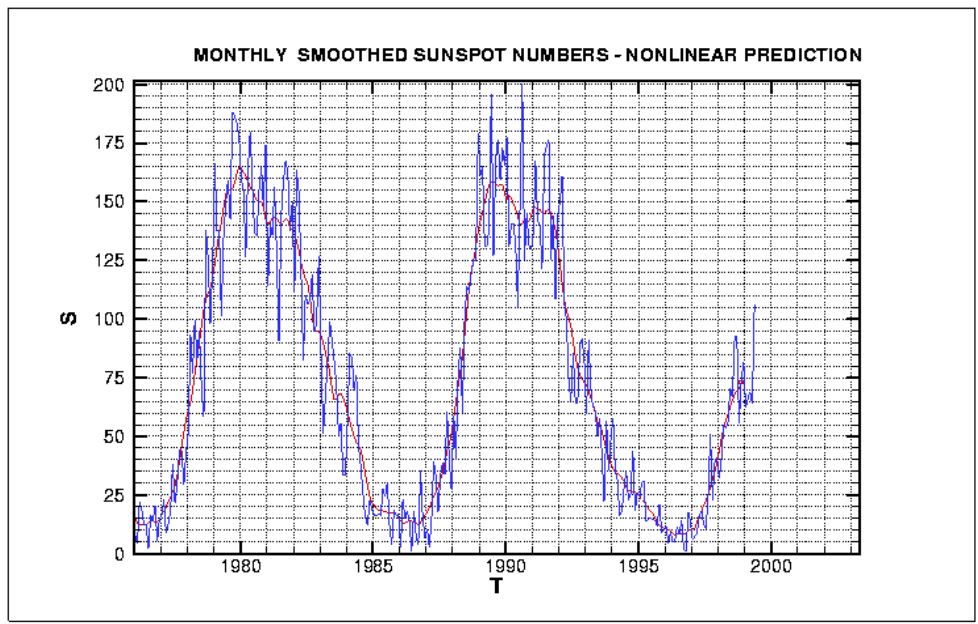

Figure 1

In order to obtain accurate predictions it is required to analyze the data recorded for long time. Figure 2 shows the whole time series of the monthly mean sunspot numbers for the period 1749.5-1998.872.

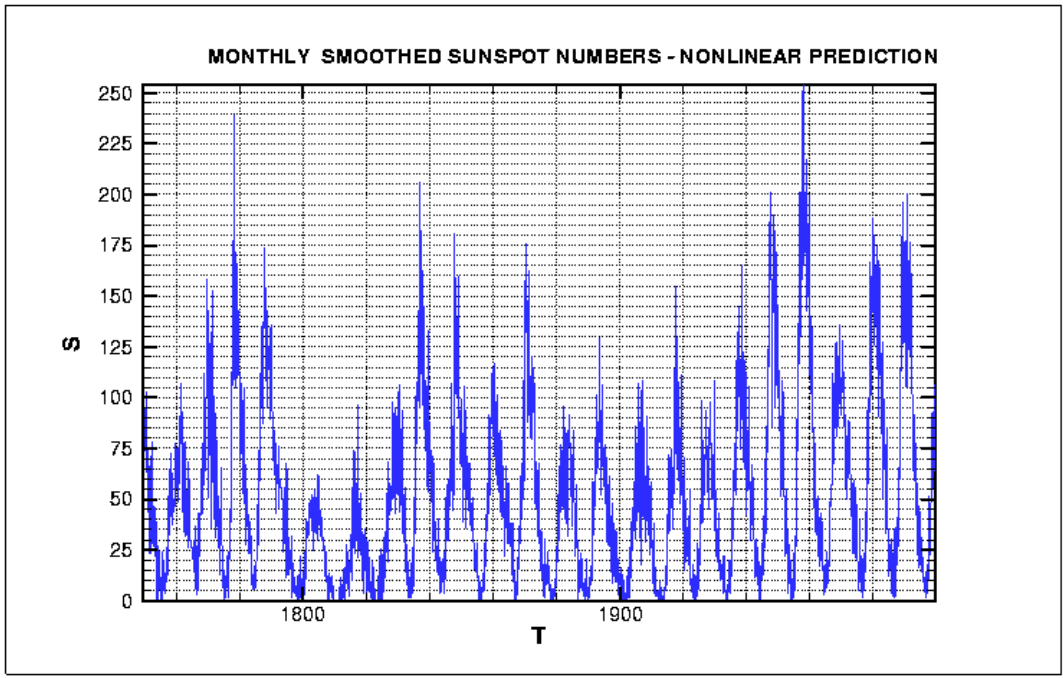

Figure 2

The intrinsic complexity in the behaviour of the sunspot numbers, suggested the possibility of a nonlinear (chaotic) dynamics governing the related process, as well pointed out by many previous works. In particular here we refer to the recent paper of Zhang [7] in which we proposed an interesting and promising nonlinear prediction method for the smoothed monthly sunspot numbers. The aim of the present paper is to support the nonlinear approach given in [7], adding a more complete and refined analysis with different nonlinear dynamics tools.

NONLINEAR DYNAMICS APPROACH

The nonlinear feature of the monthly mean sunspot number time series was not evident in the past as well documented by many different works. As example in the paper of Price, Prichard and Hogenson in 1992 [8] we founded no evidence of the presence of low dimensional deterministic behaviour in the set of the monthly mean sunspot numbers, suggesting that the filtering techniques, used to derive smoothed time series, can give some spurious evidence for the presence of deterministic nonlinear behaviour. Conversely, more recent works clearly showed strong evidences for the presence of a deterministic nonlinear dynamics governing the sunspot numbers [9],[10],[11]. Recently Kugiumtzis investigates some properties of standard and a refined surrogate technique of Prichard and Theiler to test the nonlinearity in a real time series, showing that for the annual sunspot numbers there is a strong evidence that a nonlinear dynamics is in fact present, enforcing also the idea that the sunspot numbers are in first approximation proportional to the squared magnetic field strength. [12] In the present work we used the method of surrogate data combined with the computation of linear and nonlinear redundancies, to show that the monthly mean sunspot number data contain true nonlinear dependencies [13] [14].
The use of the information-theoretic functionals, called redundancies, has at least three important advantages in comparison to other linear and nonlinear correlation analyses:

1) Various types of the redundancies can be constructed in order to test very specific types of dependence between/among variables;

2) The redundancies can be naturally evaluated as functions of time lags, so that dependence structures under study are not evaluated statically, but with respect to dynamics of a system under investigation;

3) For any type of the redundancy its linear form exists, which is sensitive to linear dependence only. These linear redundancies are used for testing quality of surrogate data in order to avoid spurious detection of nonlinearity.

The basic idea in the surrogate data correlation analysis is to compute a linear and nonlinear statistic from data under study (original) and an ensemble of realizations of a linear stochastic process (surrogates) which mimics linear properties only of the original data. If the computed statistic for the original data is significantly different from the values obtained for the surrogate set, one can infer that the data were not generated by a linear process; otherwise the null hypothesis, that a linear model fully explains the data is accepted and the data can be usefully analyzed and characterized by using well-developed linear methods.

Here we consider the nonlinear R(X,Y) redundancy of the type:

$$R(X,Y) = H(X) + H(Y) - H(X,Y)$$
$$H(X) = -\sum_{x \in X} p(x) \log p(x)$$
$$H(X,Y) = -\sum_{x \in X} \sum_{y \in Y} p(x,y) \log p(x,y)$$

where X and Y are random variables with a probability function p(x)=Pr(X=x), H(X) is the entropy and H(X,Y) is the joint entropy. Here: Y=X(t+$\tau$) and R=R($\tau$).
If the variables X and Y have zero means, unit variances and correlation matrix C, the linear redundancy L(X,Y) is of the form:

$$L(X,Y) = -\frac{1}{2} \sum_{i=1}^{2} \log(\sigma_i)$$

where $\sigma_i$ are the eigenvalues of the 2x2 correlation matrix C.

We define the test statistic as the difference between the redundancy obtained for the original data and the mean redundancy of a set of surrogates, in the number of standard deviations (SD) of the latter. Both the redundancies and redundancy based statistic are function of the time lags $\tau$. The general redundancies R detect all the dependencies contained in the data under study, while the linear redundancies are sensitive to linear structures only. Fig.3 shows the results of the computation of linear redundancy L, and nonlinear redundancy R for both the original time series of the monthly mean sunspot numbers and the related surrogate ensemble (30 realizations) as functions of time lags. We computed linear and nonlinear redundancies for 30 realizations of the surrogate time series which mimic the linear properties of the original data. We show the mean redundancies computed for the surrogate ensembles: the linear redundancy curve coincides with the linear redundancy of the original time series; whereas the general redundancy is well distinct from the general redundancy of the original data. Fig.4 shows the quantitative analysis of the differences between the redundancies. The linear redundancy L for the data and for the surrogates coincide because there is no significant difference in the linear statistic (differences <1 SD) i.e. the surrogates mimic well the linear dependences of the series, and should not be a source of spurious results in the nonlinear test. On the other hand the result for the general redundancy for the original data is clearly different from the mean redundancy for the surrogates, and the nonlinear statistic indicates highly significant differences (>2 SD). Thus the linear stochastic null hypothesis is rejected and, considering also the results from linear statistic, significant nonlinearity is detected in a reliable way on the time series.

Figure 3-4

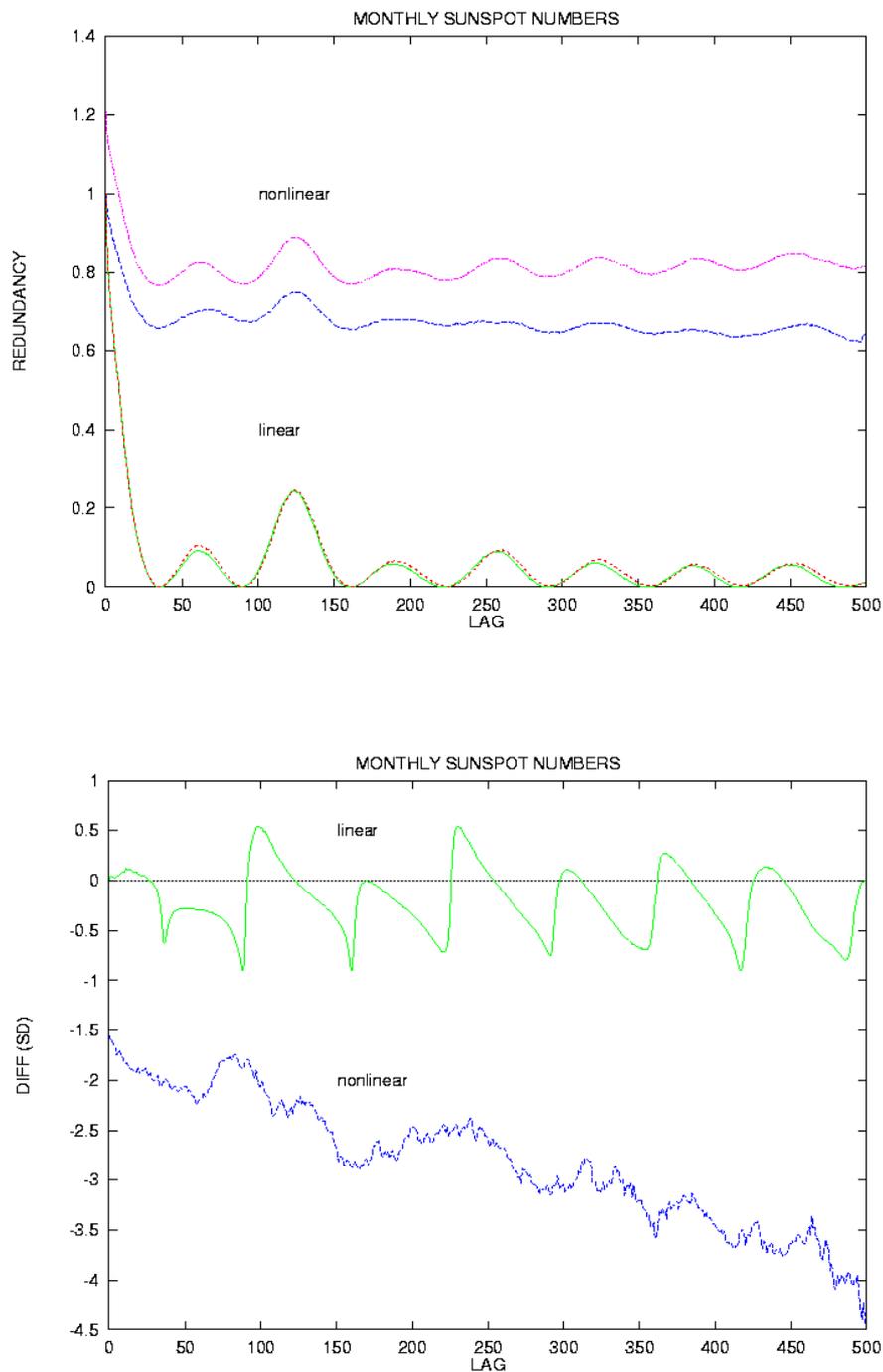

The nonlinearity analysis on the monthly mean sunspot numbers clearly supports the use of the nonlinar dynamics approach as possible prediction method. Previous preliminary works on the subject show many characteristics of the intrinsic nonlinear dynamics governing sunspot numbers. For example, Ostryakov and Usoskin in 1990 estimated their fractal dimension for different periods founding a value around 4. More recently Zhang in 1995 estimated more precisely the fractal dimension, D=2.8 ± 0.1, and the largest Lyapunov exponent, $\lambda$=0.023 ± 0.004 bits/month for the monthly mean sunspot numbers for the period 1850-1992 using the methods given by Grassberger and Procaccia and Wolf. [15], [16]. The result is the existence of a upper limit of the time scale for reliable deterministic prediction: 3.6±0.6 years. The important indication is that long-term deterministic behavior is unpredictable. Many authors proposed nonlinear prediction techniques of chaotic time series as an inverse problem for short-term prediction, with different levels of accuracy. Recently, Zhang proposed a prediction technique which improves medium-term prediction for the smoothed monthly sunspot numbers using a given local linear map to solve the inverse problem [7].

The common basis of the above works is the construction of the embedding space from the observed data which is the natural vector space in the nonlinear dynamics method. We note that also modern approaches based on neural networks prediction are based on the embedding space reconstruction in order to set some of the characteristic parameters of the model [17],[18].

Typical experimental situations concern only a single scalar time series; while the related physical systems possess many essential degrees of freedom. The powerfulness of nonlinear dynamics methods, rely on the reconstruction of the whole system's trajectory in an *"embedding space"* using the method of delay-time. The reliability of computations, performed on the reconstructed trajectory, is guaranteed by a notable theorem by Takens and Mañé (1981) [19].

Let a continuous scalar signal x(t), here the monthly mean sunspot numbers, be measured at discrete time intervals, $T_s$ (or dt), to yield a single scalar time series:

$$\{ x(t_0), x(t_0+T_s), x(t_0+2T_s), ..., x(t_0+NT_s) \}.$$

We assume that x(t) be one of the n possible state variables which completely describe our dynamical process. For practical applications, n is unknown and x(t) is the only measured information about the system. We suppose, however, that the real trajectory lies on a d-dimensional attractor in its phase space, where: d≤n. Packard et Al. and Takens have shown that starting from the time series it is possible to *"embed"* or reconstruct a *"pseudo-trajectory"* in an m-dimensional embedding space through the vectors (embedding vectors):

$$\underline{y}_1 = (\ x(t_0), x(t_0 + \tau), ..., x(t_0 + (m-1)\tau)\ )^T$$
$$\underline{y}_2 = (\ x(t_0 + l), x(t_0 + l + \tau), ..., x(t_0 + l + (m-1)\tau)\ )^T$$
$$\ldots$$
$$\underline{y}_s = (\ x(t_0 + (s-1)l), x(t_0 + (s-1)l + \tau), ..., x(t_0 + (s-1)l + (m-1)\tau)\ )^T.$$

Here $\tau$ is called *"delay-time"* or lag, and l is the sampling interval between the first components of adjacent vectors. A selection of proper values of parameters in the embedding procedure is a matter of extreme importance for the reliability of results, as well pointed out in many works [20], [21], [22], [23]. The delay time, $\tau$, for example, is introduced because in an experiment the sampling interval is in general chosen without an accurate prior knowledge of characteristic time scales involved in the process.

Takens formal criterion tells us how embedding dimension m and attractor dimension d must be related to choose a proper embedding, i.e. with equivalent topological properties:

$m \geq 2d + 1.$

Fortunately, for practical applications, this statement generally results too conservative and thus it is adequate and correct a reconstruction of attractor in a space with a lower dimensionality. Here we used a reliable method to estimate the minimum necessary embedding dimension introduced by Kennel and Abarbanel in 1994 and based on the false neighbors [24]. The idea is to eliminate "illegal projection" finding for each embedding vector, the nearest neighbor in different embedding dimensions. If the distance between the vectors in higher dimensions is very large, then we have a false nearest neighbor caused by improper embedding. When the fraction of false nearest neighbor is lesser than some threshold we are able to find the minimum embedding dimension. For the details of the method we refer to [24].

Figure 5 shows the results of the false neighbor method for the monthly mean sunspot numbers.

As clearly indicated the minimum embedding dimension value is m=5. This result is coherent with previous analyses (Zhang,1996), indicating that this time series is related to a low dimension nonlinear deterministic system described by a finite number of parameters, or by vectors in a 5 dimensional phase space. In [7] using the Grassberger-Procaccia method we founded a saturation for correlation dimension d=2.8 at m=7; on the other hand the prediction technique is based on the value m=3.

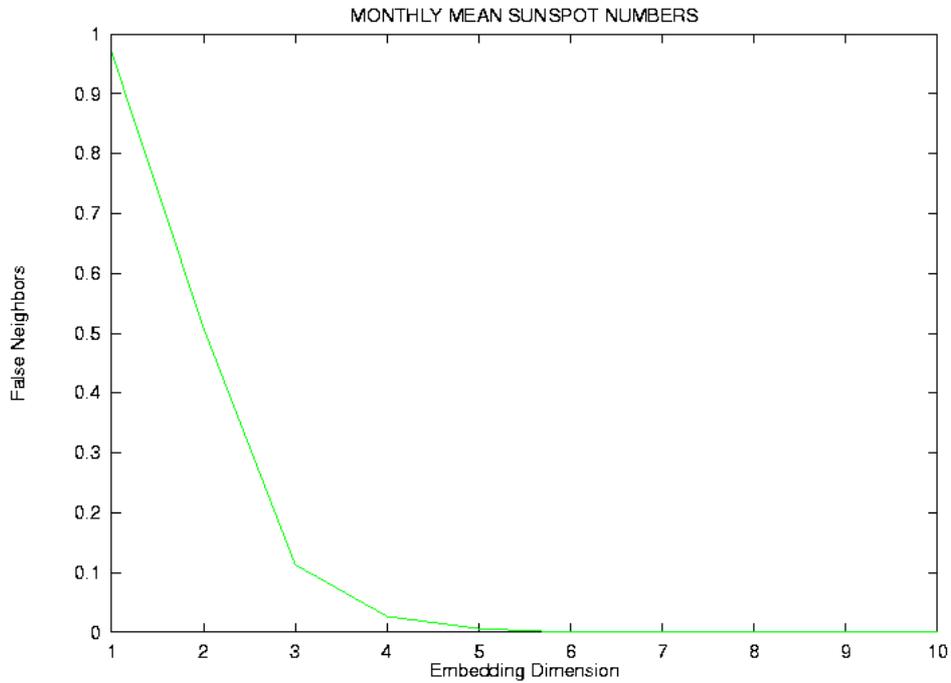

Figure 5

Here the proper choice of delay time is based on the *mutual information* of Fraser and Swinney [22], which is more adequate than autocorrelation function when nonlinear dependencies are present:

$$I(X,Y) = H(X) + H(Y) - H(X,Y)$$

Figure 6 shows the result of the computation of the mutual information for the monthly mean sunspot numbers. As we note, the first local minimum of I(X,Y) is positioned at 40dt corresponding to an interval of 3.32 years. The components of the embedding vectors can be considered independent at least with this lag.

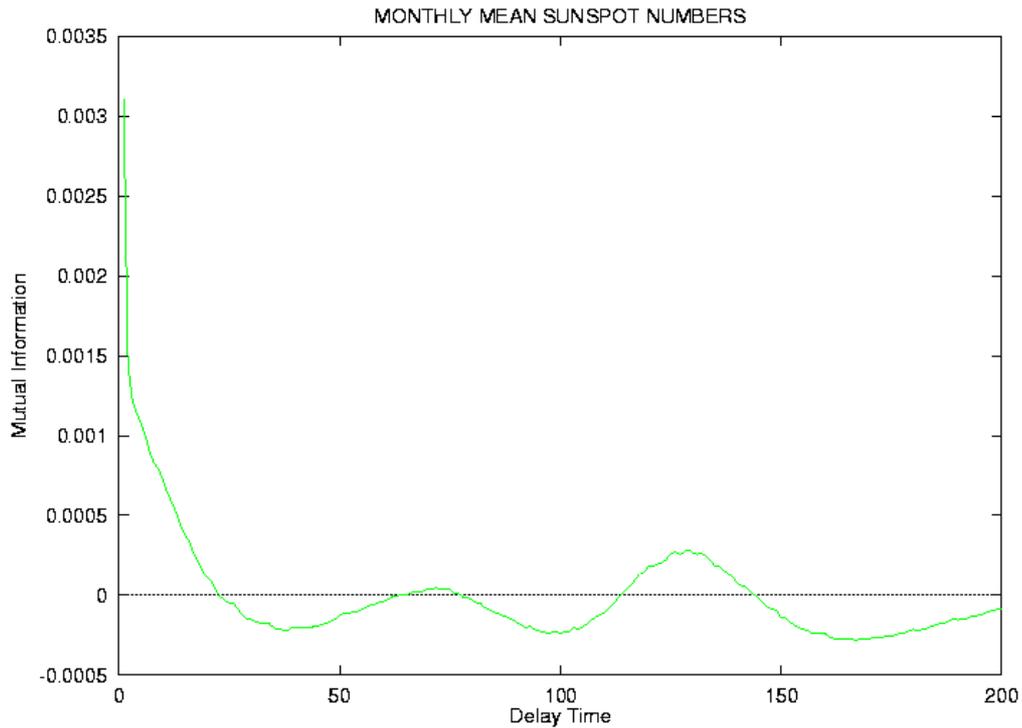

Figure 6

For a comparison, in [7] the computation of the autocorrelation function for the period 1850 -1994 gives a lag equal to 35.

Methods of nonlinear dynamics can be strongly limited by typical features of experimental situations. Correlation dimension techniques, in particular, are based on assumptions that cannot be rigorously fulfilled by experiments, especially due to the presence of broadband noise. In real cases can happens that the presence of noise results as a severe pitfall for correlation dimension algorithms, compromising the reliability of distinction between stochastic and deterministic behaviour.

Besides correlation dimension estimates, the *spectrum of Lyapunov's exponents* provides an important quantitative measure of the sensitivity to initial conditions, and moreover, from a theoretical viewpoint, it is the most useful dynamical diagnostic tool for deterministic chaotic behaviour. If the Lyapunov's spectrum contains at least one positive exponent, then the related system is defined to be chaotic and, more important, the value of this exponent yields the magnitude of predictability time scale. Furthermore, if we are able to compute the full Lyapunov-exponent spectrum, the Kolmogorov-Sinai entropy can be estimated using the Kaplan-Yorke conjecture [20]. However, as well known, there are many difficulties implied in the reliable estimation of Lyapunov's spectrum from complex experimental data [25]. This task represents a current active research area and many authors have given important improvements. Here we used a combined method deriving from works by Sano and Sawada (1985) [26]; Zeng, Eykholt and Pielke (1991) [27]; Brown, Bryant and Abarbanel (1991) [28].

The Lyapunov's exponents that come out of this procedure, based also on the phase space reconstruction, we will identify as $\lambda_i$, arranged in decreasing order: $\lambda_1 \geq \lambda_2 \geq \lambda_3 \geq ...$

Using the concepts of *local* and *global* dimensions, generally defined on the basis of previous correlation dimension computations, we determine an appropriate cut-off value for the number of exponents which can be related to the Lyapunov's dimension. In fact, following the connection postulated by Kaplan and Yorke we compute the Lyapunov's dimension by:

$$D_L = k + \frac{\sum_{i=1}^{k} \lambda_i}{|\lambda_{k+1}|},$$

where k is the maximum number of exponents that can be added before the sum becomes negative. The dimension $D_L$ is determined by only the first k+1 exponents; thus the dimension does not depend on exponents beyond the (k+1)th, which are somewhat spurious.

For the computation of the complete Lyapunov's spectrum, we selected as local dimension $d_L=5$, while the optimal value for global dimension was: $d_G=2d_L+1=11$, for the monthly mean sunspot numbers. In Figure 7 we display the results of Lyapunov's spectrum computations using the above characteristic parameters of the embedding reconstruction procedure. As we can easily verify, the "relaxation" of Lyapunov's exponents is sufficient to extrapolate quite reliable estimates from the Lyapunov's spectra. We note that, theorically, one of the exponents must be zero (in our case $\lambda_2$). More precisely, for the monthly mean sunspot numbers the single positive exponent was: $\lambda_1=0.146$ yrs$^{-1}$ suggesting a limit for reliable deterministic predictions (Lyapunov's time).

The sum of all the positive Lyapunov's exponents gives an estimation of the Kolmogorov's entropy and its inverse, multiplied by log2, gives the error doubling predictability time, tp. Thus, in our case, the estimated error-doubling predictability time gives: tp=4.72 years (56 dt). This time is the practical limit for reliable predictability.

For comparison in [7], based on the first Lyapunov's exponent with the method of Wolf et al. [16], the limit for deterministic prediction is estimated about 3.6 years.

Lyapunov dimension estimation, based on Kaplan-Yorke conjecture, gives: $D_L \approx 4.36$. The above results, unlike the correlation dimension analysis showed in [7] (d=2.8), indicate clearly an higher degree of geometrical complexity in the phase space for the monthly mean sunspot numbers.

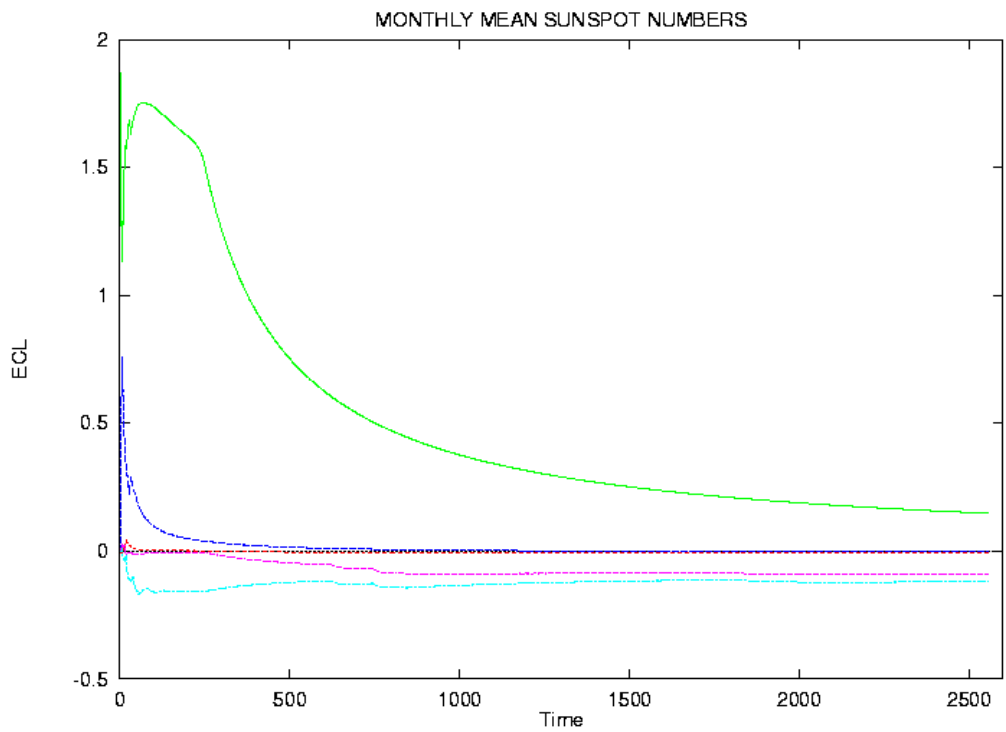

Figure 7

## SOLAR PREDICTIONS

The above complete characterization of the nonlinear dynamics governing the monthly mean sunspot numbers, allows to construct a predictive model based on the nonlinear deterministic behaviour of the embedding vectors. Here we follow essentially the approach indicated in [7] to define a smooth map for the related inverse problem. More precisely the nonlinear deterministic behaviour in the embedding space implies the existence of a smooth map $f^T$ satisfying the relation:

$$f^T(\underline{y}_t) = \underline{y}_{t+T}$$

for a given embedding vector $\underline{y}$. The inverse problem consists in the computation of this smooth map, given a time series $\{x(t)\}$, $t=1,\ldots n$. This map is the basis for the predictive model. Following the approach given in [7] we first divided the known time series into two parts: the first one: $\{x(t)\}$, $t=1,\ldots,n'$ is used to set up the smooth map $f^T$, and the other part: $\{x(t)\}$, $t=n'+1,\ldots,n$ is used to check the accuracy of the prediction model. From the above analysis we set $n'=n-t_p/dt$. In order to calculate the unknown smooth function $f^T$ we assume a local linear hypothesis for the evolution of the embedding vectors, and this is quite reliable for $T=1$. Given the last embedding vector, we select the first k neighboring vectors near the reference vector in the $m=5$ embedding space, using a distance function. Then we assume that the evolution of the selected vector is correlated with the evolution of the neighboring vectors and the parameters of this correlation are computed with the solution of a proper least squares problem in the embedding space. More precisely, the order of the matrix of the least squares problem is (kxm+1), and the predicted one step ahead vector is given by solving the least squares problem for each component of the related k neighboring vectors:

$$y'^{(i)}_{t+1} = \alpha_0 + \sum_{j=1}^{m} \alpha_j y^{(j)}_t$$

for $i=1,\ldots,m$.

This procedure is iterated for all the successive n-n' embedding vectors and the accuracy of the prediction model is evaluated by the computation of the global average predictive error:

$$<E^2>(f,k) = \frac{1}{n-n'} \sum_{t=n'+1}^{n} (\underline{y}_t' - \underline{y}_t)^2 / \sigma^2$$

The optimal model corresponds to the minimum value of $<E^2>(f,k)$ as a function of k. The whole analysis is performed for each new value added to the known part of the time series. The distribution of the optimal k values for the prediction of monthly mean sunspot numbers in the interval limited by tp is shown in Figure 8.

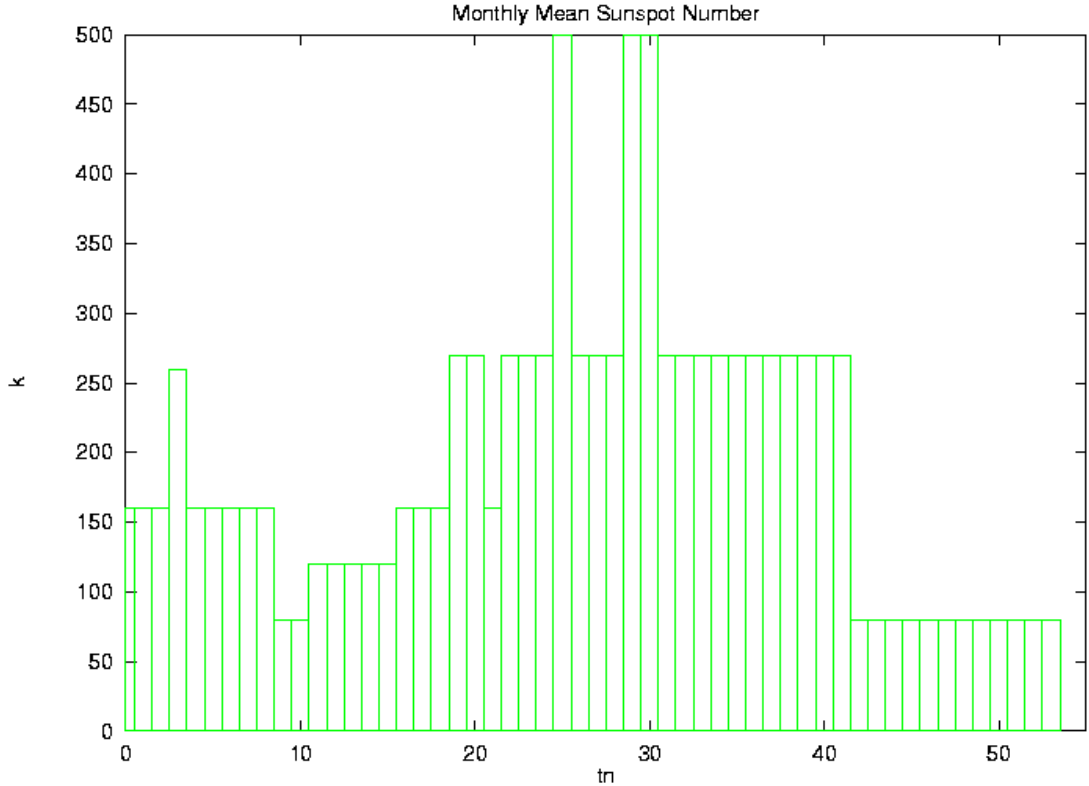

Figure 8

The original time series used in the analysis is the monthly mean sunspot data derived from SIDC archive [6] (2999 values) for the period: 1749.5- June 1999 (Figure 2). The final prediction is related to the smooth series of the smoothed monthly sunspot data, derived from the following relation:

$$\tilde{S}_n = \frac{1}{12}\left[\sum_{k=n-5}^{n+5} S_k + \frac{1}{2}(S_{n+6} + S_{n-6})\right]$$

where $S_k$ is the mean value of S for the month k. This choice is motivated by the fact that even if the monthly mean sunspot series contains high level of broadband noise which can degrades severely the accuracy of the predictions, smoothing is not an invariant process in dynamical systems and may affects some intrinsic features of the original data [13].
In Figures 9,10 we show the results of the nonlinear prediction model for a period limited by the error doubling predictability time tp.

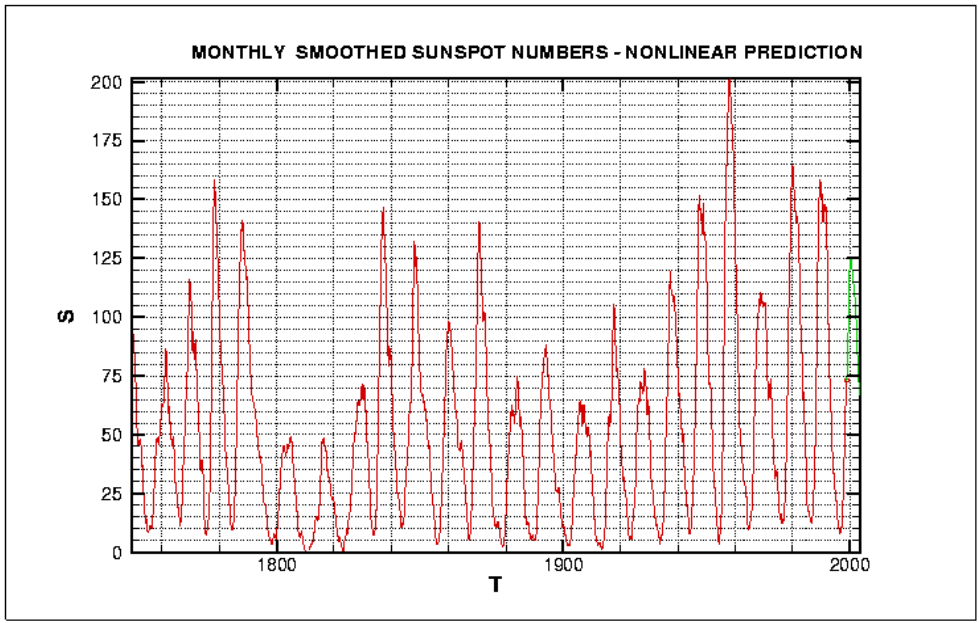

Figures 9,10

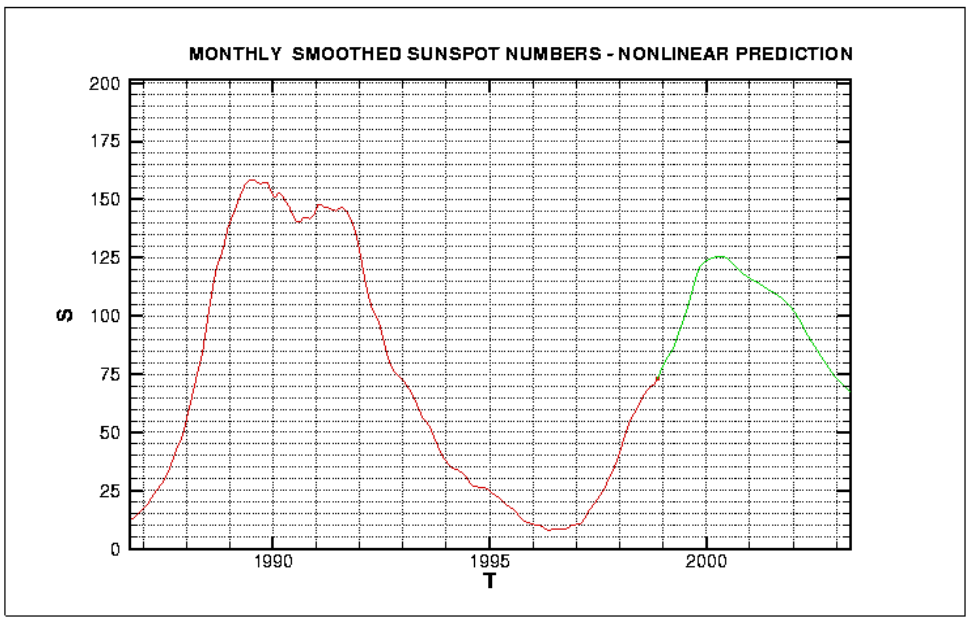

The red solid line is the known smoothed monthly sunspot series and the green solid line is the corresponding predicted behaviour covering the period: 1998.79, 2003.26. The red symbols are the observed values derived after May 1999. As we can see the maximum of the smoothed monthly sunspot numbers for the 23th cycle is predicted at 2000.28 with the value 125.6. Based on this prediction the value is comparable with the maximum reached in 1937.5 (113.5).

To compare, a posteriori, the accuracies of predictions obtained using the most efficient methods proposed in literature, as example, we show in Figure 11 the predictions given by SIDC (June 1999).

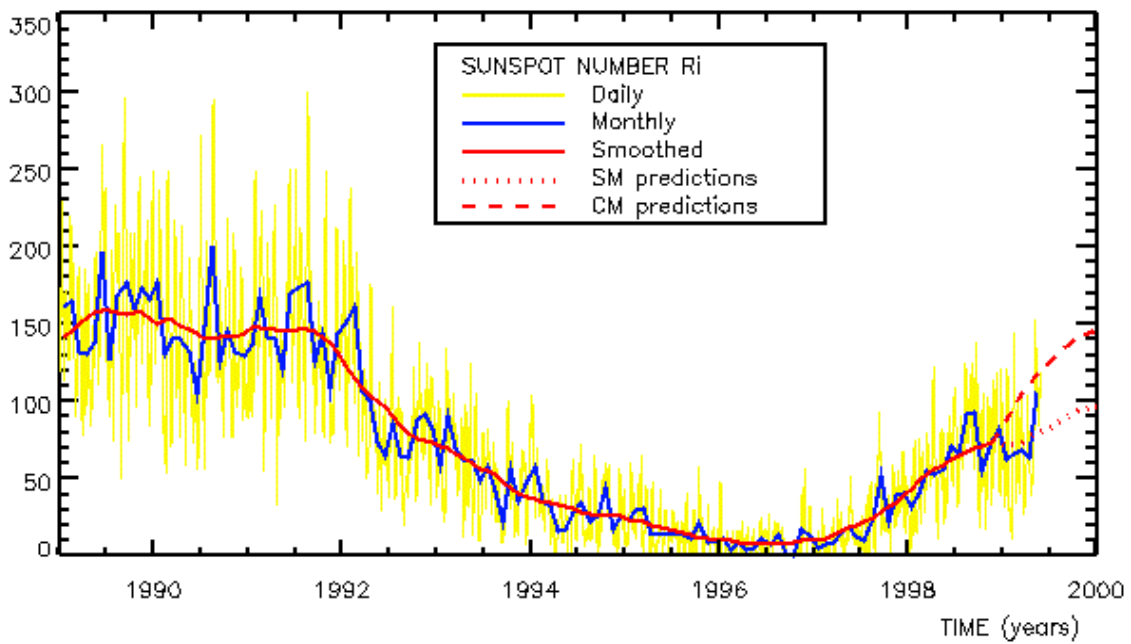

Figure 11 (SIDC)

SM red dots is a classical prediction method based on an interpolation of Waldmeier's standard curves, and CM red dashed is a combined method (due to K. Denkmayr) a non-parametric regression technique coupling a dynamo-based estimator with Waldmeier's idea of standard curves [29].
Typical precursor methods, geomagnetic and solar, predict high amplitudes with maximum values about 160 at April-May 2000 [30],[31]. (Figure 12)

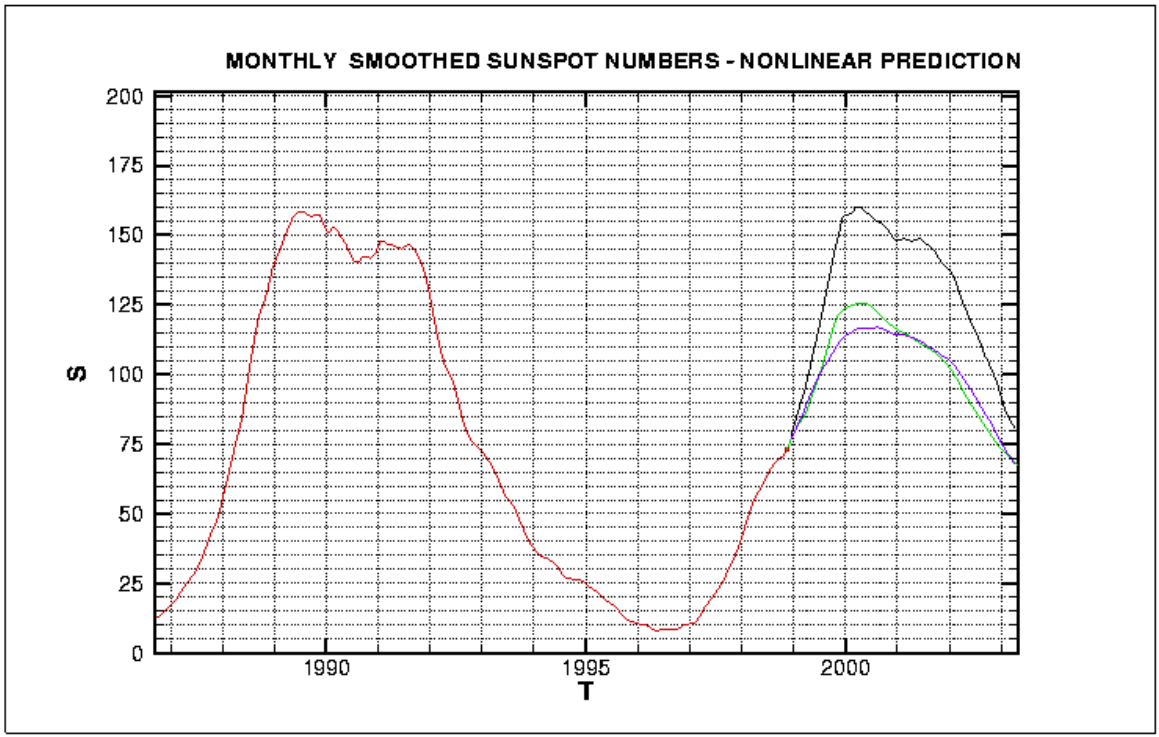

Figure 12

Black solid line is the prediction from a precursor method based on solar and geomagnetic activity (IPS) [32]. Blue solid line is the prediction from the method of A.G. McNish and J.V. Lincoln and modified using regression coefficients and mean cycle values computed for Cycles 8 through 20 (SIDC). It is important to point out the coherence of these methods to predict the phase of the next maximum. The global evaluation of the accuracy of predictions, for the 23th solar cycle, is postponed to the complete recording of the observed data.

CONCLUSIONS

The problem of prediction of smoothed monthly sunspot numbers is examined, with particular attention to the nonlinear dynamics approach. The intrinsic complexity of the related time series strongly affects the accuracy of the phase and amplitude predictions.
Starting from a previous recent work, we checked the reliability of a forecasting model based on concepts of nonlinear dynamics theory applied to experimental time series, such as embedding phase space, Lyapunov spectrum, chaotic behaviour. The analysis clearly pointed out the nonlinear-chaotic nature with limited predictability of the monthly mean sunspot time series as suggested in many previous preliminary works. The model is based on a local hypothesis of the behaviour on the embedding space, utilizing an optimal number k of neighbour vectors to predict the future evolution of the current point with the set of characteristic parameters determined by several previous parametric computations. The performances of this method suggest its valuable insertion in the set of the called statistical-numerical prediction techniques, like Fourier analyses, curve fitting, neural networks, climatological, etc. The main task is to set up and to compare, using the data for the current 23th solar cycle, this promising numerical nonlinear prediction technique, essentially based on an inverse problem, with the most accurate predictive methods, like the so-called "precursor methods", which appear now reasonably accurate in predicting "long-term" Sun activity.